\documentclass[aps,prl,amsmath,amssymb,reprint,superscriptaddress]{revtex4-1}
\usepackage{graphicx} 
\usepackage[version=4]{mhchem}

\newcommand{\D}{\mathrm{d}} % upright differentials

\begin{document}

\title{Magnetoentropic signatures of skyrmionic phase behavior in FeGe}

\author{Joshua D. Bocarsly}
\affiliation{Materials Department and  Materials Research Laboratory, 
University of California, Santa Barbara, California 93106}
%\affiliation{Materials Department, 
%University of California, Santa Barbara, California 93106} 

\author{Ryan F. Need}
\affiliation{Materials Department and Materials Research Laboratory, 
University of California, Santa Barbara, California 93106}

\author{Ram Seshadri}
\affiliation{Materials Department and Materials Research Laboratory, 
University of California, Santa Barbara, California 93106}
%\affiliation{Materials Department, 
%University of California, Santa Barbara, California 93106} 
%\affiliation{Department of Chemistry and Biochemistry, 
%University of California, Santa Barbara, California 93106}

\author{Stephen D. Wilson}
\email{stephendwilson@ucsb.edu}
\affiliation{Materials Department and Materials Research Laboratory, 
University of California, Santa Barbara, California 93106} 
%\affiliation{Materials Research Laboratory, 
%University of California, Santa Barbara, California 93106}

\date{\today}
\begin{abstract}
We demonstrate that magnetocaloric measurements can rapidly reveal details of the phase diagrams of high temperature 
skyrmion hosts, concurrently yielding quantitative latent heats of the field-driven magnetic phase transitions. 
Our approach addresses an outstanding issue in the phase diagram of the skyrmion host FeGe by showing that 
DC magnetic anomalies can be explained in terms of entropic signatures consistent with a phase diagram containing a single pocket of skyrmionic order and 
a Brazovskii transition.
\end{abstract}

\maketitle

\begin{figure}[t!]
	\centering
	\includegraphics[width=\columnwidth]{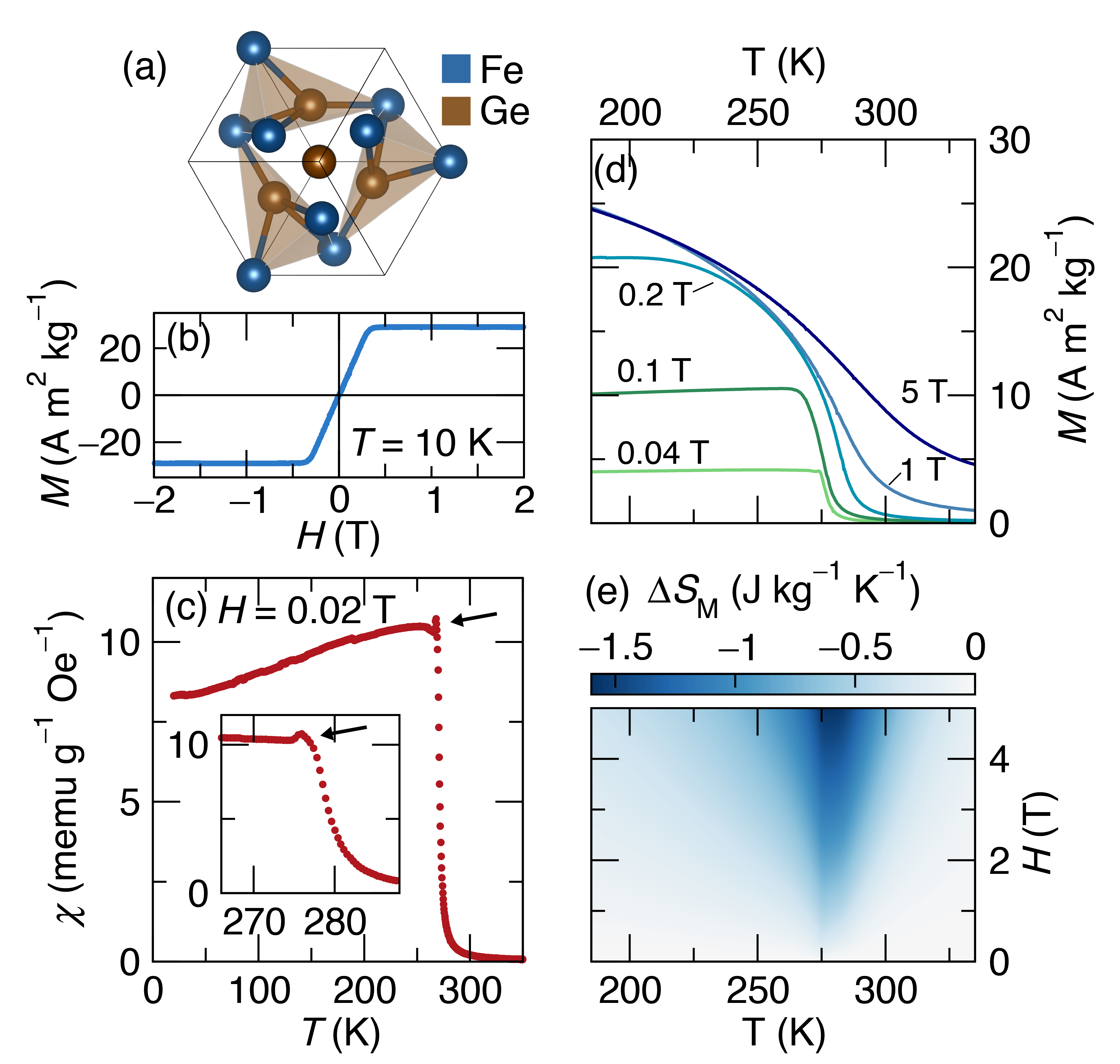}
	\caption{(a) Crystal structure of cubic B20 FeGe (spacegroup $P2_13$), shown along the (111) axis. (b) Magnetization as a function of field collected at 10\,K is very sharp, saturates at low field, and shows no hysteresis. (c) Magnetization as a function of temperature collected under applied field $ H$\,=\,20\,mT shows an anomaly near $T_C$. (d) $M(T)$ collected under different applied fields. This is a subset of the data set (18 total fields) used to calculate the course-grained map of $\Delta S_M (H, T)$ shown in (e).} 
	\label{fig:mag}
\end{figure}

Magnetic anomalies corresponding to skyrmion lattice ordering or ``precursor'' states are well known in chiral 
helimagnets such as MnSi and FeGe \cite{Ludgren1970, Haraldson1972, Ishikawa1976, Bak1980c, Ishikawa1984, Lebech1989, Bogdanov1989, Bogdanov1994} and were observed long before the first reciprocal 
space \cite{Muhlbauer2009, Pappas2009} or real space \cite{Yu2010} observations of magnetic skyrmions. In general, 
these anomalies appear as subtle bumps and kinks in the magnetization expected for a ferromagnet near its 
magnetic transition temperature, as illustrated in Fig.\,\ref{fig:mag}. In skyrmion hosts, these features represent 
magnetization steps expected for the first-order phase transitions between topologically distinct spin states. In real materials, however
these discontinuities are always smeared out by experimental convolution and inherent 
thermal/configurational disorder. This often renders mapping the bulk magnetic phase diagrams of skyrmion hosts a 
subtle endeavor, and discrepancies have arisen regarding the number of distinct topological phases that exist in 
key materials \cite{Wilhelm2011, Wilhelm2012b, Cevey2013, Moskvin2013, Bauer2012, Bauer2016}.

This problem is exacerbated in high temperature skyrmion hosts, where direct calorimetric techniques identifying 
topological phase boundaries (\emph{e.g.} heat capacity studies) suffer from large lattice background signals. 
The B20 high temperature skyrmion host FeGe is a prominent example of this challenge, where several reports suggest 
that the skyrmion \emph{A} phase in FeGe is in fact broken into several sections, each hosting distinct skyrmionic 
states \cite{Wilhelm2011, Wilhelm2012b, Cevey2013, Moskvin2013}.  The inability to directly quantify the entropic response from each of these phases in FeGe hearkens to 
parallel studies of the low temperature skyrmion host MnSi, where similar multiple ``\emph{A}-phase'' 
states were proposed \cite{Kadowaki1982} but eventually precluded \emph{via} high resolution heat capacity 
measurements \cite{Bauer2012, Bauer2013}.  Resolving whether there is only a single pocket in the ``\emph{A} phase'' 
that hosts skyrmionic spin texture or multiple in FeGe remains an open question.

More broadly, the continued unveiling of magnetic skyrmions in materials near and above room temperature and their 
potential uses in practical applications \cite{Vinet2011, Nagaosa2013, Tokunaga2015, Yu2011, Yu2012, Emori2013, Woo2016, Stolt2017} has further highlighted the need 
to quantify the thermodynamically distinct spin states in their high temperature magnetic phase diagrams. New materials continue to be 
discovered, many with near-room-temperature skyrmion states \cite{Stolt2018, Tokunaga2015, Karube2016, Nayak2017, Phatak2016, Hou2017}. Precise and quantitative techniques for rapidly 
interpreting magnetic anomalies in this new realm of materials and for ultimately surveying thermodynamically distinct magnetic states in their phase diagrams are needed. 

Here we present a rapid DC magnetization technique for mapping the magnetocaloric response of skyrmion 
hosts.  This method is effective even at high temperatures and is sensitive to the field driven entropy changes associated 
with entering/exiting the first-order phase boundaries expected for topologically distinct spin states. As a result, 
the magnetic phase diagram for a given compound can be mapped in under 24 hours, and the entropy 
changes associated with a given state can be quantified. We leverage this technique to address an outstanding issue 
in the high temperature skyrmion material FeGe by demonstrating that the entropic response can be understood \emph{via}
a single skyrmion ``\emph{A}-phase'' and a nearby line of first-order phase transitions representing Brazovskii transitions 
into a fluctuation disordered state. 

\begin{figure}[t]
	\centering
	\includegraphics[width=\columnwidth]{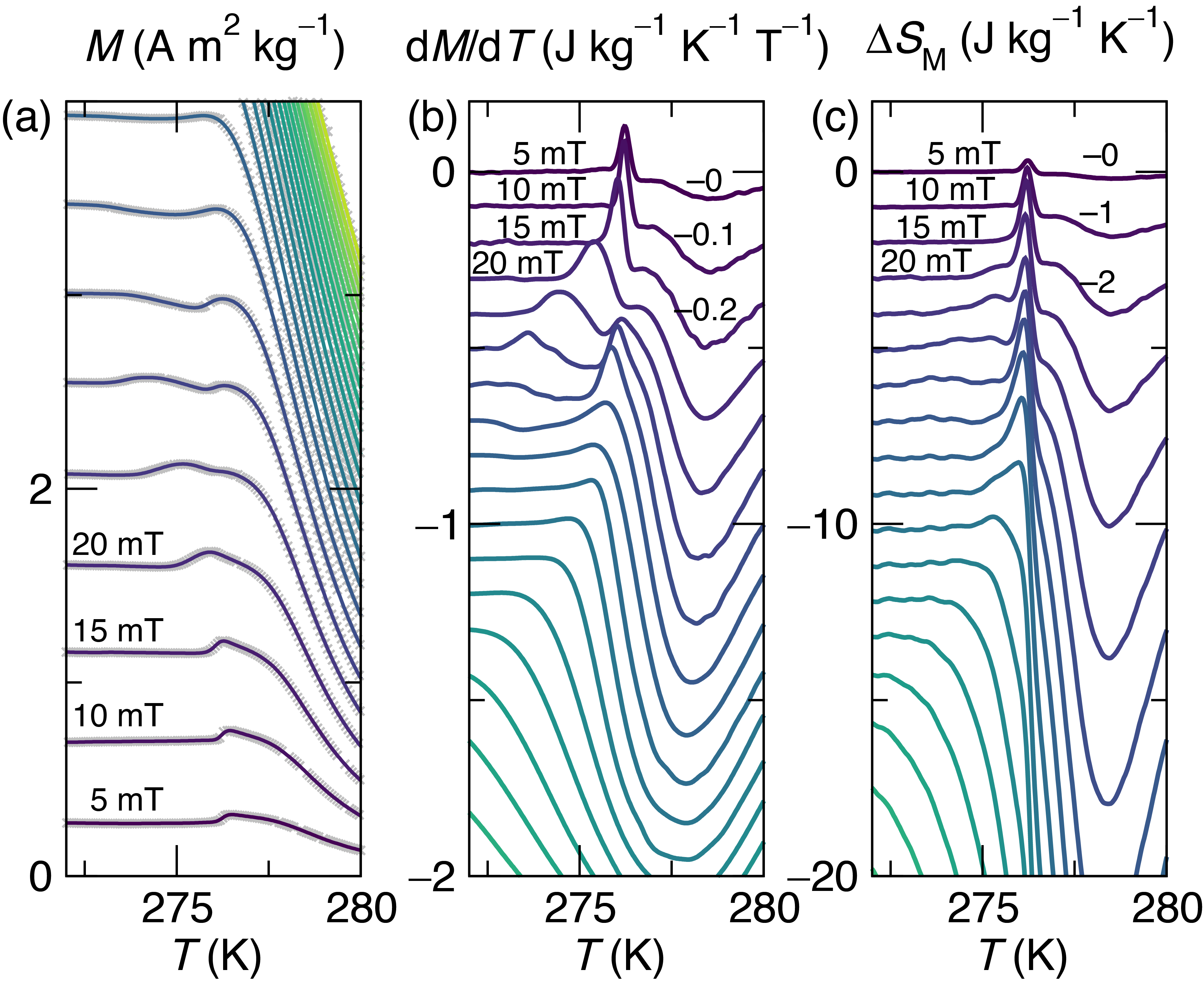}
	\caption{The process for obtaining high-resolution magnetoentropic information using Eq.\,\ref{eqn:deltaSm}. (a) DC $M(T)$ data taken at many closely-spaced fields (24 fields between 5 and 120 mT) (b) Temperature derivatives of magnetization $\D M/\D T = \D S/\D H$ are calculated directly using Tikhonov regularization. For visual clarity, the curves are each offset by 0.1\,J\,kg$^{-1}$\,K$^{-1}$\,T$^{-1}$. The antiderivatives of the calculated derivatives are shown as colored lines in (a), and match the raw data (grey crosses) very well. (c) Integrals of the $\D M/ \D T$ curves with respect to field give the isothermal magnetic entropy change at each temperature and applied field. Curves are each offset by 1\,J\,kg$^{-1}$\,K$^{-1}$.}
	\label{fig:lines}
\end{figure}

Magnetocaloric effects can be quantified as the magnitude of isothermal entropy change upon magnetization, 
$\Delta S_M (H,T)$, near a magnetic phase transition. $\Delta S_M (H,T)$ is obtained from the Maxwell relation 
$\left(\D S/\D H\right)_T =\left(\D M/\D T\right)_H$, where $S$ is the total 
entropy, $H$ is the magnetic field, $M$ is the magnetization, and $T$ is the temperature. This allows the isothermal 
entropy change upon application of field $H$ to be calculated from bulk DC magnetic measurements at many fields and 
temperatures using

\begin{equation}
\Delta S_M(T,H) = \int_{0}^{H}{\left(\frac{\D M}{\D T}\right)_{H^\prime} dH^\prime}
\label{eqn:deltaSm}
\end{equation}

\noindent Comparisons to heat capacity measurements carried out under field have validated the use of
this approach, even for the analysis of first-order phase transitions if suitable measurement parameters are chosen \cite{Caron2009,Porcari2012,Balli2009}. Measuring $M(T)$ under different applied magnetic fields
and calculating $\D M/ \D T$ allows a map of $\Delta S_M(T,H)$ to be obtained using Eq.\,\ref{eqn:deltaSm}. 

To date, applications of these methods have been largely limited to using DC magnetization to calculate $\Delta S_M(T, H)$ at a few temperatures and fields
to evaluate materials for applications in magnetic refrigeration \cite{Franco2012} and to determine critical constants \cite{Xu2017}. For these applications, low data densities and simple numerical methods are adequate.
However, in order to apply these techniques to measure, in resolution, the entropic effects of the subtle field-driven phase transitions in magnetic skyrmion hosts,
far higher data densities are required and more sophisticated data processing is needed to separate signal from noise.

To demonstrate this concept, single crystals of the high temperature skyrmion host FeGe were grown using a standard iodine vapor transport 
technique (see Supplemental Material \cite{supp}) and a Quantum Design DynaCool Vibrating Sample Magnetometer (VSM) was used to collect two
datasets: a ``course-grained'' set taken while sweeping temperature at a rate of 7\,K\,min$^{-1}$ with fields ranging from 20\,mT to 5\,T  and a ``fine-grained'' set taken while sweeping at a rate of 1\,K\,min$^{-1}$ at closely-spaced fields around the
magnetic transition.  The former was taken to evaluate the general high field magnetocaloric response and the latter to analyze the skyrmion phase transition. By operating the VSM continuously, tens of thousands of data points are collected in an $\approx$18 hour measurement span. 
The $\D M/ \D T$ numerical derivatives cannot be calculated using traditional finite differences without introducing unnacceptable noise.
Rather, a statistical technique based on Tikhonov regularization \cite{Stickel2010} was employed. Briefly, the derivatives are determined so as to simultaneously
minimize the deviation of their antiderivatives from the data and the roughness. From these smooth derivatives, the integrals with respect to field were evaluated to obtain $\Delta S_M(T,H)$. Details of the technique are included in the Supplemental Material \cite{supp}. 

\begin{figure*}[t]
	\centering
	\includegraphics[width=1.0\textwidth]{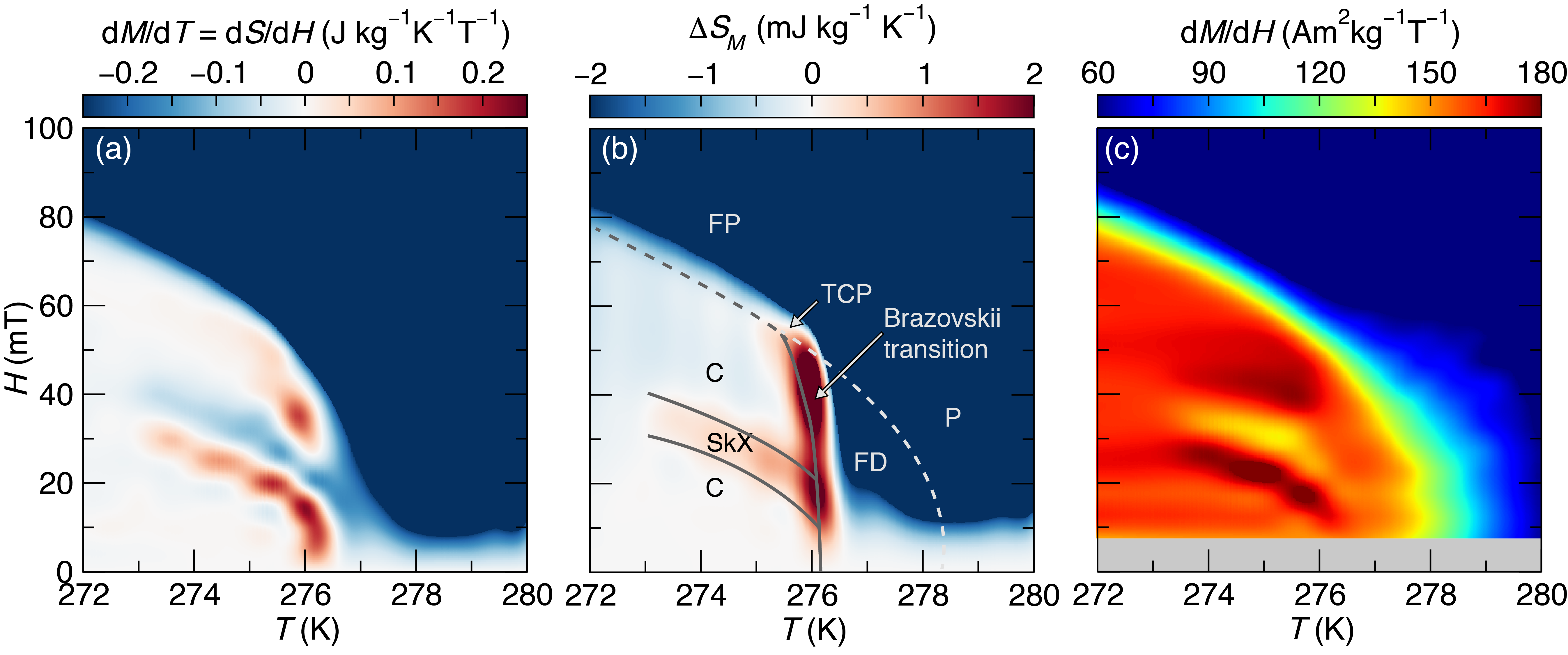}
	\caption{Detailed magnetoentropic maps of FeGe near the ordering temperature. (a) A map of $\D{M}/ \D T = \D S/ \D H$ reveals clear ridges (red) and valleys (blue) indicating lines of first-order phase transitions. Note that the ridges and valleys are actually continuous; the segmented appearance is an artifact of the 2-D interpolation. (b) Map of $ \Delta S_M(T,H)$. (c) $\D M/\D H$ calculated from the DC magnetization dataset. (a) is used to find the phase boundaries of the phase diagram drawn on (b), where solid lines represent first-order phase transitions.The dashed line between C and P indicates a continuous transition. The dashed line between FD and FP represents a crossover. P: paramagnetic, FD: fluctuation disordered, FP: field polarized, C: conical, SkX: skyrmion lattice, TCP: tricritical point.}
	\label{fig:maps}
\end{figure*}

Figure~\ref{fig:mag}(a) shows the B20 magnetic lattice of FeGe looking along the (111) axis of the cubic unit cell. A chiral spin state is known to manifest below 280 K in this system with the helix propagating along this (111) wave vector and moments rotating in the plane orthogonal to this axis.  Upon applying a modest field, this helical spin state rapidly tilts into a conical phase and eventually into a polarized ferromagnet state as shown in Fig.~\ref{fig:mag}(b). The low field susceptibility $\chi(T)$ is shown in Fig.~\ref{fig:mag}(c) and the characteristic cusp near $T_C$ is apparent.  Magnetization data at higher fields are shown in Fig.~\ref{fig:mag}(d) while the resulting $\Delta S_M$ determined from this course-grained sampling of the phase diagram is shown in Fig.~\ref{fig:mag}(e).  As expected, a negative peak in $\Delta S_M$ is seen near the magnetic ordering temperature as the magnetic field aligns paramagnetic spins and decreases the entropy of the system. 

At lower fields, however, the magnetization and magnetocaloric behavior are more complex.  Figure~\ref{fig:lines}(a) illustrates how the low field magnetization evolves as a function of temperature under a series of closely spaced fields near the magnetic ordering temperature. This rich behavior is then processed into $\D M / \D T$ at each temperature and field point as shown in Fig.~\ref{fig:lines}(b). The final integrated $\Delta S_M$ curves are plotted in Fig.~\ref{fig:lines}(c). These data are then presented in Fig.\,\ref{fig:maps} as $(T, H)$ maps of $\D M / \D T$,  $\Delta S_M$, and instantaneous DC susceptibility $\D M / \D H$ near the onset of the ``\emph{A} phase" cusp. 

$\D M/\D T = \D S/\D H$ can be viewed as a thermodynamic capacity which gives complementary information to traditional measurements of heat capacity $C = T(\D S/ \D T)$.  Peaks and valleys in $\D S/\D H$ can indicate field-driven first-order phase transitions and ultimately can give entropies of transitions. In the map shown in Fig.~\ref{fig:maps}(a), the high-field region is blue, indicating the conventional (negative) magnetocaloric behavior of a ferromagnet discussed above. At lower fields and temperatures, however, a white region ($\D S/\D H \approx 0$) can be seen with clear ridges (red lines) and valleys (blue lines) corresponding to phase transformations within that region. 

When integrated over field ($\Delta S_M$), the phase regions separated by features in $\D S/\D H$ are visualized in terms of their entropy, as seen in Fig.~\ref{fig:maps}(b). The sharp nearly-vertical phase line near 276\,K denotes a line of first-order phase transitions between the ordered state and the fluctuation disordered state, as discussed later. At temperatures below this first-order line, a single, small pocket of increased entropy (about 0.3\,J\,kg$^{-1}$\,K$^{-1}$) is observed about the expected skyrmion phase. All other points in the white region, which corresponds to the ordered helical and conical phases, can be reached without a change in entropy from the zero-field state.  The observation that the skyrmion lattice shows distinctly higher entropy than the conical phase is consistent with the idea that the skyrmion lattice is stabilized by thermal fluctuations. As further reference, Fig.\,\ref{fig:maps}(c) shows a map of static $\D M/\D H$ illustrating the onset of an enhanced susceptibility at $\approx 279$\,K, far above the first-order line and indicative of the onset of the fluctuation disordered regime. Anomalies in the susceptibility map of Fig.\,\ref{fig:maps}(c) bracket both the upper and lower field phase boundaries of the single ``\emph{A} phase" skyrmion state resolved in the $\Delta S_M$ map. 

The assignment of a skyrmion lattice pocket approximately 3\,K in width and 10\,mT in height within the conical phase is consistent with previous phase diagrams of FeGe based on AC and DC susceptibility, specific heat, and small angle neutron scattering measurements. \cite{Wilhelm2011, Wilhelm2012b, Cevey2013, Moskvin2013} However, variations in AC susceptibility and SANS neutron scattering intensities caused speculation that the conventional skyrmion state, termed the $A_1$ pocket, was neighbored by between one and three additional ``$A$ phase'' pockets. Notably none of the signatures of these new ``$A$ phase'' pockets arise from thermodynamic measurements nor via the identification of broken symmetries, and here, our thermodynamic magnetoentropic measurements resolve that none of those regions except the expected main $A$-phase show increased entropy relative to the helimagnetic state. Therefore, we conclude that the previous signatures of additional states near the ``$A$ phase'' arise from dissipative processes or mixed phase regions due to the nearby line of first order Brazovskii transitions. Any true thermodynamic phases must have much smaller skyrmion numbers than the skyrmion lattice phase and entropies nearly indistinguishable from the topologically trivial helical and conical phases.

\begin{figure}[t]
	\centering
	\includegraphics[width=\columnwidth]{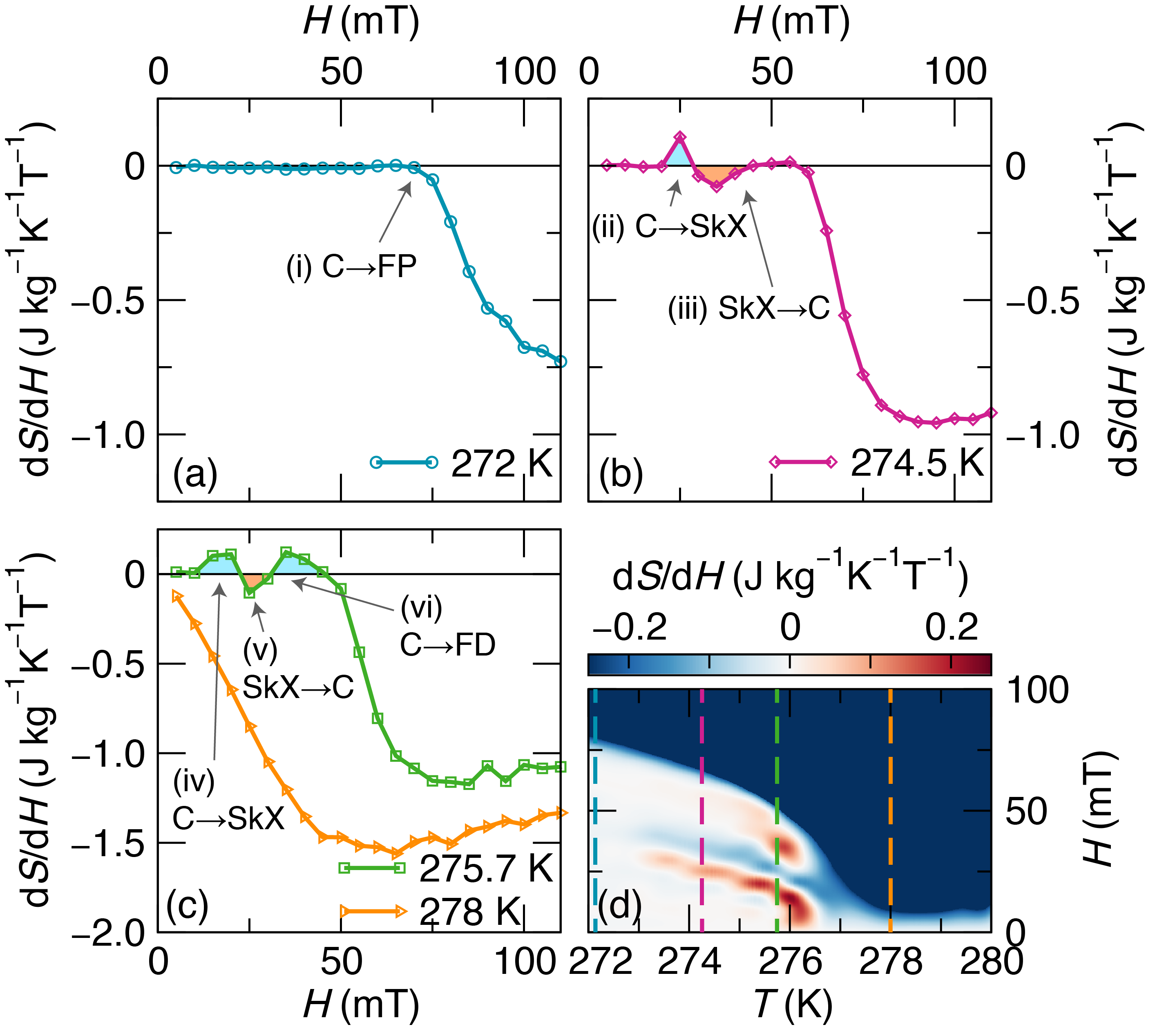}
	\caption{(a-c) shows $\D S/\D H$ \emph{vs.} $H$ at four representative temperatures. Field-driven phase transitions are easily found: first-order transitions show up as peaks (ii-vi) in the this thermodynamic capacity, while continuous phase transitions show up as changes in the slope (i). Integration of the peaks gives entropies of transitions and latent heats, 
as shown in Table\,\ref{tbl:latent}.  (d) gives a reproduction of the heatmap of $\D S/\D H$ \emph{vs.} $T$ and $H$, as shown in Figure~\ref{fig:maps}b with the
 slices shown in (a-c) overlaid as dashed colored vertical lines. Refer to the caption of Fig.~\ref{fig:maps} for the definitions of the phase abbreviations.
}
	\label{fig:latent}
\end{figure}

To further quantify the entropies associated with the phase boundaries in Fig.~\ref{fig:maps}(a), Fig.~\ref{fig:latent}(a) shows $\D S/\D H$ \emph{vs.} $H$ cuts at fixed temperatures across the phase diagram of FeGe. At temperatures below the skyrmion lattice phase (Fig.~\ref{fig:latent}(a)), the conical to field polarized phase transition can be seen as a sudden change in slope of the $\D S/\D H$ \emph{vs.} $H$ curve. At all fields below this critical field, it can be seen that $\D S/\D H$ is zero. This indicates that there is no change in entropy as the system is polarized from the helical magnetic state, through the conical state until the collinear ferromagnetic state is reached. Once in the ferromagnetic state, application of a magnetic field suppresses spin fluctuations, reducing entropy as expected. One consequence of this constant entropy in the low field phase, is that there is no signature in $\D S/\D H$ for the helical to conical phase transition at low field.

\begin{table}[]
	\centering
	\caption{Latent entropies and heats of transitions as determined by integrating the $\D S/\D H$ curves shown in Fig.~\ref{fig:latent}. The errors are a generous estimate based on performing the integration at several closely-spaced temperatures.
	}
	\label{tbl:latent}
	\begin{tabular}{@{}llrr@{}}
		\toprule
		& transition        & $\Delta S$ (mJ kg$^{-1}$ K$^{-1}$) & $Q$ (mJ kg$^{-1}$) \\ \colrule
		(i)   & C $\rightarrow$ FP  &   \emph{na}      &   \emph{na}      \\
		(ii)  & C $\rightarrow$ SkX & 0.25(5)     & 69(14)      \\
		(iii) & SkX $\rightarrow$ C & $-$0.35(5)  & $-$96(14)  \\
		(iv)  & C $\rightarrow$ SkX & 0.9(1)     & 248(28)     \\
		(v)   & SkX $\rightarrow$ C & $-$0.29(2)  & $-$80(6)   \\
		(vi)  & C $\rightarrow$ FD  & 0.81(3)      & 223(8)     \\ \botrule
	\end{tabular}
\end{table}

Turning to Fig. 4 (b), $\D S/\D H$ cuts along $H$ near 274.5\,K show there is both a peak and a valley prior to entering the field polarized state. Hence as field is increased, there is first an absorption of heat and then a release of heat. This is consistent with the expected entropic signature of first-order phase transitions into and out of the skyrmion lattice phase based on heap capacity measurements of low-temperature skyrmion hosts \cite{Bauer2013,Cevey2013,Bauer2016}. These peak and valley features form the extended ridges in ($H$,$T$) space (Fig.\,\ref{fig:maps}(b)) that define the top and bottom of the skyrmion lattice phase. 

At higher temperature ($T\approx$ 276 K), the nearly vertical ridge in the $\D S/\D H$ is split into a lower and an upper section by the intersection of the skyrmion phase boundaries (Fig.\,\ref{fig:latent}(d)). This vertical ridge indicates another line of first-order phase transitions where the application of a magnetic field disorders the system. This is consistent with the theory of a Brazovskii scenario of a strong fluctuations driving the magnetic ordering into a line of first-order transitions terminating in a tricritical point at  nonzero field (here, around 50 mT)  \cite{Brazovskii1975, Brazovskii1976, Janoschek2013}. Crucially, because the slope of this ridge in $(T,H)$ space is negative, application of a field drives the system from the ordered helimagnetic state to the fluctuation disordered state: hence the sign of the $\D S/\D H$ is positive. Therefore, this unique transition appears as a striking line of anomalous (positive) $\D S/\D H$ on the magnetocaloric maps. The entropies associated with crossing each of these phase boundaries are summarized in Table\,\ref{tbl:latent}.

This global picture shows that the very complex shape of the DC magnetic anomalies in FeGe can in fact be elegantly associated with the magnetoentropic response expected for a phase diagram containing a single thermodynamic $A$ phase (skyrmion lattice) contained within the conical phase that borders out of a line of first-order Brazovskii transitions. To verify that features of this phase diagram were not affected by the use of several single crystals, the same procedure was carried out on a fixed single crystal ($\approx 0.1$\,mg) and yielded the same phase diagram (Supplemental Material Fig.\,S3 \cite{supp}). This is consistent with observations of very low anisotropy fields in FeGe \cite{Haraldson1972}. 

In summary, we have demonstrated a rapid magnetoentropic mapping technique that harnesses DC magnetization data to resolve the magnetic entropies associated with the complex phase diagrams of helimagnets in very high resolution. This technique allows for the clear demarcation of thermodynamic phase boundaries in FeGe, which have been difficult to study in traditional calorimetry measurements due to a high ordering temperature and accompanying large lattice background. We observe clear entropic signatures of transitions into and out of a single skyrmion lattice phase as well as observe a nearly vertical line of first-order transitions terminating in a tricritical point, consistent with the first-order Brazovskii transition observed in MnSi. The technique presented here is expected to be of significant utility for the rapid discovery and study of new skyrmion hosts, especially those with transitions near and above room temperature.

\begin{acknowledgements}
This work was supported by  the National Science Foundation through the MRSEC Program of the National Science 
Foundation through DMR-1720256 (IRG-1). J.D.B. and R.F.N are supported by NSF Graduate Research Fellowship Program 
under Grant No. 1650114 and Grant No. 1144085, respectively. 
\end{acknowledgements}

\end{document}